October 94

On the connection between the pinch technique and the background field method.

(To appear in Physical Review D)

Joannis Papavassiliou

Department of Physics, New York University, 4 Washington Place,

New York, NY 10003, USA.

ABSTRACT

The connection between the pinch technique and the background field method is further explored. We show by explicit calculations that the application of the pinch technique in the framework of the background field method gives rise to exactly the same results as in the linear renormalizable gauges. The general method for extending the pinch technique to the case of Green's functions with off-shell fermions as incoming particles is presented. As an example, the one-loop gauge independent quark self-energy is constructed. We briefly discuss the possibility that the gluonic Green's functions, obtained by either method, correspond to physical quantities.

The pinch technique (PT) [1] is an algorithm that allows the construction of modified gauge independent (g.i.) $n$-point functions, through the order by order rearrangement of Feynman graphs contributing to a certain physical and therefore ostensibly g.i. process, such as an S-matrix element (Fig.1) or a Wilson loop. Its original motivation was to devise a consistent truncation scheme for the Schwinger-Dyson equations that govern the dynamics of gauge theories. In fact, it has been extensively employed as a part of a non-perturbative approach to continuum QCD [2]. On the other hand, most of its recent applications have been in the area of electroweak physics [3-5]. The simplest example that demonstrates how the PT works is the gluon two point function (propagator). Consider the S-matrix element $T$ for the elastic scattering of two fermions of masses $m_1$ and $m_2$. To any order in perturbation theory $T$ is independent of the gauge fixing parameter $\xi$. On the other hand, as an explicit calculation shows, the conventionally defined proper self-energy (collectively depicted in graph 1(a) depends in on $\xi$. At the one loop level this dependence is canceled by contributions from other graphs, such as 1(b), 1(c), and 1(d), which, at first glance, do not seem to be propagator-like. That this cancellation must occur and can be employed to define a g.i. self-energy, is evident from the decomposition:

$$T(s, t, m_1, m_2) = T_1(t, \xi) + T_2(t, m_1, m_2, \xi) + T_3(s, t, m_1, m_2, \xi) , \qquad (1)$$

where the function $T_1(t)$ depends only on the Mandelstam variable $t = -(\hat{p}_1 - p_1)^2 = -q^2$, and not on $s = (p_1 + p_2)^2$ or on the external masses. Typically, self-energy, vertex, and box diagrams contribute to $T_1$, $T_2$, and $T_3$, respectively. Such contributions are $\xi$ dependent, in general. However, as the sum $T(s, t, m_1, m_2)$ is g.i., it is easy to show that Eq. (1) can be recast in the form

$$T(s, t, m_1, m_2) = \hat{T}_1(t) + \hat{T}_2(t, m_1, m_2) + \hat{T}_3(s, t, m_1, m_2) , \qquad (2)$$

where the $\hat{T}_i$ ($i = 1, 2, 3$) are *individually* $\xi$-independent. The propagator-like parts of graphs, such as 1(e), 1(f), and 1(g), which enforce the gauge independence of $T_1(t)$, are called "pinch parts". They emerge every time a gluon propagator or an elementary three-gluon vertex contribute a longitudinal $k_\mu$ to the original graph's numerator. The action of such a term is to trigger an elementary Ward identity of the form $\slashed{k} = (\slashed{p}+\slashed{k}-m)-(\slashed{p}-m)$ once it gets contracted with a $\gamma$ matrix. The first term removes the internal fermion propagator (that is a "pinch"), whereas the second vanishes on shell. The g.i. function $\hat{T}_1$ is identified with the contribution of the new effective propagator. Its one-loop closed form, renormalized in the $\overline{MS}$ scheme, reads

$$\hat{\Pi}_{\mu\nu}(q) = t_{\mu\nu}\hat{\Pi}(q) \qquad (3)$$

$$\hat{\Pi}(q) = -q^2 bg^2 [\ln(\frac{-q^2}{\mu^2}) - \frac{67}{33}] \ . \qquad (4)$$

$t^{\mu\nu} = (g^{\mu\nu} - \frac{q^\mu q^\nu}{q^2})$, $b = \frac{11 c_a}{48 \pi^2}$ is the coefficient in front of $-g^3$ in the usual one loop $\beta$ function, and $c_a$ the Casimir operator for the adjoint representation. [$c_a = N$ for $SU(N)$]

This procedure can be generalized to an arbitrary $n$-point function. In particular, the g.i. three and four point functions $\hat{\Gamma}_{\mu\nu\alpha}$ and $\hat{\Gamma}_{\mu\nu\alpha\beta}$ derived in [6] and [7] respectively, satisfy the following Ward identities [8]:

$$\begin{aligned}q_1^\mu \hat{\Gamma}_{\mu\nu\alpha}(q_1,q_2,q_3) &= t_{\nu\alpha}(q_2)\hat{d}^{-1}(q_2) - t_{\nu\alpha}(q_3)\hat{d}^{-1}(q_3) \\ q_1^\mu \hat{\Gamma}_{\mu\nu\alpha\beta}^{abcd} &= f_{abp}\hat{\Gamma}_{\nu\alpha\beta}^{cdp}(q_1+q_2,q_3,q_4) + c.p. \ ,\end{aligned} \qquad (5)$$

where $\hat{d}^{-1}(q) = [q^2 - \hat{\Pi}(q)]$, the $f^{abc}$ are the structure constants of the gauge group, and "c.p." stands for "cyclic permutations". The above results have been originally obtained in the non-covariant light-cone gauge [2], and later in the linear renormalizable $R_\xi$ gauges [6]. The purpose of this paper is to discuss recent important developments in this field.

Recently, an important connection between the PT and the background field method (BFM) [9] has been established [10-12]. In particular, it was shown that when QCD is quantized in the context of BFM, the conventional $n$-point functions, calculated with the BFM Feynman rules, *coincide* with those obtained via the PT, for the special value $\xi_Q = 1$ of the gauge fixing parameter $\xi_Q$, used to gauge-fix the "quantum" field. For any other value of $\xi_Q$ the resulting expressions differ from those obtained via the PT. However, the BFM $n$-point functions, for any choice of $\xi_Q$, satisfy exactly the same Ward identities as the PT $n$-point functions (Eq. (5) for example). Based on these observations, it was argued [10] that the PT is but a special case of the BFM, and represent one out of an infinite number of equivalent choices, parameterized by the values chosen for $\xi_Q$. Alternatively, one could say that the Feynman gauge ($\xi_Q = 1$) in the BFM has the special property of rendering pinching trivial; thus, it provides an alternative, more economical way, for obtaining the PT results. It is important to emphasize however that the aforementioned equivalence between the PT and the BFM has only been established for specific, one-loop examples (two, three, and four point functions), mainly due to the fact that no formal understanding of the PT algorithm exists thus far. Although we have no progress to report in this direction, in the present paper we explore additional issues related to the connection between the PT and the BFM. In particular, we show via explicit one-loop calculations that:

(a) The PT, when applied in the context of the BFM, for any value of the gauge-fixing parameter $\xi_Q$, gives exactly the same answer as in any other gauge checked so far. This exercise furnishes an additional check for the internal consistency of the PT.

(b) After extending the PT to the fermionic sector, we show that the g.i. quark-propagator obtained, coincides with the expression obtained for the quark propagator in the context of the BFM, again for the special value of $\xi_Q = 1$.

(c) Finally, we conjecture that the PT and the BFM $n$-point functions correspond to physical quantities, which, at least in principle, can be measured.

To the extend that the BFM $n$-point functions display a residual (even though mild) $\xi_Q$-dependence, one may still apply the PT algorithm, in order to obtain a g.i. answer. It turns out that the PT results can be recovered for *every* value of $\xi_Q$ as long as one properly identifies the relevant pinch contributions concealed in the rest of the graphs contributing to the $S$-matrix element. These contributions vanish for $\xi_Q = 1$, but are *non-vanishing* for any other value of $\xi_Q$. It seems therefore that, after the PT procedure is completed the same result emerges, regardless of the gauge fixing procedure (BFM, $R_\xi$, light-cone, etc), or the value of the gauge fixing parameter ($\xi_Q$, $\xi$, $n_\mu$, etc) used [13]. Therefore, as far as the PT is concerned, the difference between various gauge fixing procedures is only operational; in the BFM, for instance, the pinch contributions to the gluon two, three and four point functions are ultra-violet finite [14].

We now proceed to apply for the first time the PT in the context of the BFM. As shown in [10], $n$-point functions, even when computed in the framework of the BFM, depend *explicitly* on $\xi_Q$. The one-loop gluon self-energy, for example, reads:

$$\Pi_{\mu\nu} = \Pi_{\mu\nu}^{(\xi_Q=1)} + C(\xi_Q) t_{\mu\nu}$$
$$= \hat{\Pi}_{\mu\nu} + C(\xi_Q) t_{\mu\nu} \ . \tag{6}$$

$C(\xi_Q)$ does not depend on $q^2$; its explicit value is

$$C(\xi_Q) = \frac{\lambda}{4}[-8 + \lambda] c_a g^2 \ , \tag{7}$$

where $\lambda = 1 - \xi_Q$. It is amusing to notice that $C(\xi_Q)$ vanishes not only for $\xi_Q = 1$, but also for the less appealing value of $\xi_Q = -7$.

We next compute the propagator-like pinch contributions of the amplitude shown in Fig.1. using the BFM Feynman rules. It should be emphasized that no additional assumptions will be made, other than the straightforward application of the PT rules, which are common for any type of gauge-fixing procedure. The main characteristics of the Feynman rules in the BFM [9] are that the gauge fixing parameters for the "background" (classical) and the "quantum" fields are different ($\xi_C$ and $\xi_Q$ respectively [15]), the three and four-gluon vertices are $\xi_Q$-dependent at tree-level, and the couplings to the ghosts are modified (they are however $\xi_Q$-independent). In particular, the three-gluon vertex assumes the form [10](omitting a factor $if_{abc}$)

$$\Gamma_{\mu\nu\alpha}^{(0)} = (\frac{1-\xi_Q}{\xi_Q}) \Gamma_{\mu\nu\alpha}^P + \Gamma_{\mu\nu\alpha}^F \ , \tag{8}$$

with
$$\Gamma^P_{\mu\nu\alpha} = (q+k)_\nu g_{\mu\alpha} + k_\mu g_{\nu\alpha}$$
$$\Gamma^F_{\mu\nu\alpha} = 2q_\mu g_{\nu\alpha} - 2q_\nu g_{\nu\alpha} - (2k+q)_\alpha g_{\mu\nu} .$$
(9)

$\Gamma^F_{\mu\nu\alpha}$ satisfies the Ward identity $q^\alpha \Gamma^F_{\mu\nu\alpha} = [k^2 - (k+q)^2]g_{\mu\nu}$. $\Gamma^P_{\mu\nu\alpha}$ gives rise to pinch parts, when contracted with $\gamma$ matrices. Clearly, it vanishes for $\xi_Q = 1$, and so do the longitudinal parts of the gluon propagators; therefore pinching in this gauge is zero. However, for any other value of $\xi_Q$ the pinch contributions are non-vanishing.

Setting $\int_k \equiv \frac{g^2 \mu^{4-n}}{(2\pi)^n} \int d^n k$, the dimensionally regularized loop integral, we obtain from the box diagrams [Fig.1(d) and the crossed, not shown]

$$B^P_{\mu\nu} = i\lambda c_a q^4 \left[ t_{\mu\nu} \int_k \frac{1}{k^4(k+q)^2} - \frac{\lambda}{2} t_{\mu\rho} t_{\sigma\nu} \int_k \frac{k^\rho k^\sigma}{k^4(k+q)^4} \right] ,$$
(10)

and from the vertex diagrams

$$[V^P_1]_{\mu\nu} = -i\lambda c_a q^2 t_{\mu\nu} \int_k \frac{1}{k^4} ,$$
$$[V^P_2]_{\mu\nu} = i\lambda c_a q^2 \left[ t_{\mu\nu} \int_k \frac{2qk}{k^4(k+q)^2} + \frac{\lambda}{2} t_{\mu\rho} t_{\sigma\nu} \int_k \frac{k^\rho k^\sigma}{k^4(k+q)^4} \right] - B^P_{\mu\nu} - [V^P_1]_{\mu\nu} .$$
(11)

$[V^P_1]_{\mu\nu}$ originates from graph (c) of Fig.1, the self-energy corrections for the external fermions (not shown), and the mirror graphs (also not shown); $[V^P_2]_{\mu\nu}$ from graph (b) of Fig.1 and its mirror graph (not shown). Notice that $B^P_{\mu\nu} = B^P_{\mu\nu}|_{R_\xi}$. Adding the contributions of Eq. (10) and Eq. (11) we obtain the total pinch contribution to the gluon self-energy:

$$\Pi^P_{\mu\nu} = i\lambda c_a q^2 \left[ t_{\mu\nu} \int_k \frac{2qk}{k^4(k+q)^2} + \frac{\lambda}{2} q^2 t_{\mu\rho} t_{\sigma\nu} \int_k \frac{k^\rho k^\sigma}{k^4(k+q)^4} \right]$$
$$= -\frac{\lambda}{4}[-8 + \lambda] c_a g^2 t_{\mu\nu}$$
$$= -C(\xi_Q) t_{\mu\nu} .$$
(12)

The final step in constructing $\tilde{\Pi}_{\mu\nu}$, the PT gluon self-energy in the BFM, is to append the pinch contributions from Eq. (12) to the conventional expression of Eq. (6). The answer is

$$\tilde{\Pi}_{\mu\nu} = \hat{\Pi}_{\mu\nu} .$$
(13)

So, the PT self-energy in the BFM ($\widetilde{\Pi}_{\mu\nu}$) is *identical* to that constructed in the $R_\xi$ or the non-covariant light-cone gauge ($\hat{\Pi}_{\mu\nu}$).

Turning to the conventional three and four-point functions, it is straightforward to establish that, when calculated in the context of the BFM for an arbitrary value of $\xi_Q$, they are also $\xi_Q$-dependent. The answer has the general form:

$$\Gamma_{\mu\nu\alpha}(\xi_Q, q_1, q_2, q_3) = \hat{\Gamma}_{\mu\nu\alpha}(q_1, q_2, q_3) + R_{\mu\nu\alpha}(\xi_Q, q_1, q_2, q_3)$$
$$\Gamma_{\mu\nu\alpha\beta}(\xi_Q, q_1, q_2, q_3, q_4) = \hat{\Gamma}_{\mu\nu\alpha\beta}(q_1, q_2, q_3, q_4) + S_{\mu\nu\alpha\beta}(\xi_Q, q_1, q_2, q_3, q_4) \quad .$$
(14)

Both $R_{\mu\nu\alpha}$ and $S_{\mu\nu\alpha\beta}$ are ultra-violet finite, obey Bose symmetry, and vanish at $\xi_Q = 1$. In addition, as one can verify by an explicit calculation, they satisfy the following Ward identities:

$$q_1^\mu R_{\mu\nu\alpha} = C(\xi_Q)[q_3^2 t_{\nu\alpha}(q_3) - q_2^2 t_{\nu\alpha}(q_2)]$$
$$q_1^\mu S_{\mu\nu\alpha\beta}^{abcd} = f_{abp} R_{\nu\alpha\beta}^{cdp}(q_1 + q_2, q_3, q_4) + c.p.$$
(15)

This is of course expected; indeed, since both $\Gamma_{\mu\nu\alpha}(\xi_Q, q_1, q_2, q_3)$ and $\hat{\Gamma}_{\mu\nu\alpha}(q_1, q_2, q_3)$ satisfy Eq. (5), with $d \leftrightarrow \hat{d}$, and since $\Pi_{\mu\nu}$ and $\hat{\Pi}_{\mu\nu}$ are related by Eq. (6), Eq. (15) must be satisfied. The above argument provides an additional consistency test; we emphasize however that, in the context of the PT, Eq. (15) can only be verified through an explicit calculation, but cannot be established a *priori*, based on more general arguments.

One can construct $\xi_Q$-independent effective three and four gluon vertices in the BFM, $\widetilde{\Gamma}_{\mu\nu\alpha}$ and $\widetilde{\Gamma}_{\mu\nu\alpha\beta}$, respectively, following directly the PT rules. To that end one has to use the BFM Feynman rules and isolate vertex-like pieces from all relevant Feynman diagrams contributing to the appropriate scattering processes to a given order (Fig.2). The sums $\Gamma_{\mu\nu\alpha}^P$ and $\Gamma_{\mu\nu\alpha\beta}^P$ of all such vertex-like contributions satisfy $\Gamma_{\mu\nu\alpha}^P = -R_{\sigma\nu\alpha}$ and $\Gamma_{\mu\nu\alpha\beta}^P = -S_{\sigma\nu\alpha\beta}$. Adding $\Gamma_{\mu\nu\alpha}^P$ and $\Gamma_{\mu\nu\alpha\beta}^P$ to the regular $\xi_Q$-dependent contributions $\Gamma_{\mu\nu\alpha}$ and $\Gamma_{\mu\nu\alpha\beta}$, respectively, one obtains

$$\widetilde{\Gamma}_{\mu\nu\alpha} = \Gamma_{\mu\nu\alpha} + \Gamma_{\mu\nu\alpha}^P = \Gamma_{\mu\nu\alpha}|_{(\xi_Q=1)} = \hat{\Gamma}_{\mu\nu\alpha}$$
$$\widetilde{\Gamma}_{\mu\nu\alpha\beta} = \Gamma_{\mu\nu\alpha\beta} + \Gamma_{\mu\nu\alpha\beta}^P = \Gamma_{\mu\nu\alpha\beta}|_{(\xi_Q=1)} = \hat{\Gamma}_{\mu\nu\alpha\beta} \quad .$$
(16)

The PT has been so far applied to $n$-point functions, where all $n$ incoming particles are off-shell gauge bosons (gluons). It is possible however to extend the PT to the case of $n$-point functions involving fermions (quarks) as incoming off-shell particles. Such an exercise is useful, for two reasons. First, one is interested in exploring the range of applicability of the PT by itself. Second, the connection between the PT and the BFM has only been established through explicit examples; it is therefore important to determine whether or not the aforementioned connection holds in the fermionic sector as well.

In order to apply the PT in the fermionic sector, one has to embed a given $n$-point function with $N$ off-shell fermion legs ($N \leq n$) into a process containing $N$ gluons as incoming particles. A g.i. quark propagator can be extracted, for example, by applying the PT to a process such as $gluon + quark \rightarrow gluon + quark$ (Fig.3) . Similarly, the g.i. gluon-quark vertex, with all three incoming momenta off-shell may be obtained by considering a process of the form $quark + quark \rightarrow 2 \ gluons + 2 \ quarks$.

For simplicity we will treat the case of the quark propagator. It is important to notice that the conventional expression for the quark propagator is gauge-dependent both in BFM and the $R_\xi$ gauges. In fact, the two answers are identical; one can be obtained from the other by simply exchanging $\xi \leftrightarrow \xi_Q$. The gauge dependent answer is given by

$$\Sigma^{ij}(p) = \Sigma^{ij}(p)|_{(\xi_Q=1)} + \lambda g^2 c_f \delta^{ij} \left[ -(\not{p} - m) \int_k \frac{1}{k^4} + (\not{p} - m) \int_k \frac{1}{[\not{k} + \not{p} - m]k^4} \ (\not{p} - m) \right], \tag{17}$$

where $c_f$ is the Casimir eigenvalue of the quark representation. We notice that the gauge-dependent term in the r.h.s. of Eq. (17), even in the BFM context, is no longer ultra-violet finite. This is to be contrasted with the gluon $n$-point functions, which, as already mentioned, have ultra-violet finite gauge-dependent terms, at least for the $n$=2,3,4 cases, which have been explicitly calculated. The relevant pinch parts, some of which are schematically shown in Fig.3e, Fig.3g, and Fig.3i, exactly cancel the $\xi_Q$-dependent terms in the r.h.s. of Eq. (17). The technical details of how such a cancellation proceeds will be presented elsewhere. The g.i. one-loop effective quark self-energy reads:

$$\tilde{\Sigma}^{ij}(p) = \hat{\Sigma}^{ij}(p) = \Sigma(p)^{ij}|_{(\xi_Q=1)} = \Sigma(p)^{ij}|_{(\xi=1)} \ . \quad (18)$$

We see that the g.i. answer obtained from the extension of the PT to the fermionic sector (quark propagator) again coincides with the conventional expression calculated at $\xi_Q = 1$ (or $\xi = 1$ for the $R_\xi$ gauges). It would be interesting to check if the same is true for the off-shell gluon-quark vertex [16]. The results of this study will be presented in a future communication.

An important open question is if the g.i. quantities extracted via the PT (and equivalently the BFM) correspond to physical quantities. Using Eq. (4), it is straightforward to verify that, up to finite constant terms, which can be absorbed in the final normalization [17], the one-loop expression for $\hat{T}_1$ is :

$$\hat{T}_1 = \bar{u}_1 \gamma_\mu u_1 \{ \frac{g^2}{q^2[1 + bg^2 \ln(\frac{-q^2}{\mu^2})]} \} \bar{u}_2 \gamma^\mu u_2 \quad , \quad (19)$$

where $u_i$ are the external quark spinors. Thus, up to the kinematic factor $\frac{1}{q^2}$, the r.h.s. of Eq. (19) is the one-loop running coupling [18]. Equivalently, modulo finite constant terms, the expression of Eq. (19) is the Fourier transform of the static quark-antiquark potential, in the limit of very heavy quark masses [19] Clearly, the quark-antiquark potential is a physical quantity, which, at least in principle, can be extracted from experiment, or measured on the lattice. In fact, as was recently realized [20], when one computes the one-loop contribution to the scattering amplitude $q\bar{q} \to q\bar{q}$ of quarks with mass $M$, retaining leading terms in $\frac{q^2}{M^2}$, one arrives again at the expression of Eq. (19). So in principle, one can extract the quantity of Eq. (19) from a scattering process, in which the momentum transfer $q^2$ is considerably larger than the QCD mass $\Lambda^2$, so that perturbation theory will be reliable, and, at the same time, significantly smaller than the mass of the external quarks, so that the sub-leading corrections of order $O(\frac{q^2}{M^2})$ can be safely neglected. Top-quark scattering, for example, could provide a physical process, where the above requirements are simultaneously met.

These observations lead to the conjecture that, the PT (or BFM) expressions for the gluonic $n$-point functions correspond (up to finite constant terms) to the static potential of a system of $n$ heavy quarks. One obvious way of further testing this conjecture (although it would not conclusively prove it) is to determine through an explicit one-loop calculation, whether or not the PT (and BFM) expressions for the three (four) point functions are physically equivalent to the static potential of a system of three (four) heavy quarks. Calculations in this direction are already in progress. Regardless of the validity of the previous conjecture, however, it would clearly be very useful to establish a formal connection between the PT and the BFM for arbitrary Green's functions. An important step for accomplishing such a task would be the formulation of the PT at the level of the path integral (generating functional).

In conclusion, in this paper we showed that the proper application of the PT in the context of the BFM gives rise to exactly the same $n$-point function as in the context of the $R_\xi$ gauges. Thus, the calculational simplifications of the BFM, especially for the value $\xi_Q = 1$ of the gauge fixing parameter, may be freely exploited. Furthermore, the PT was applied for the first time to the case of fermion (quark) self-energies. The g.i self-energy so obtained coincides with the conventional one, again for the special value of $\xi_Q = 1$. The generalization of the arguments presented above to the electro-weak sector of the Standard Model, is technically more involved, but conceptually straightforward.

**Acknowledgments:** The author thanks K. Philippides, K. Sasaki, and M. Schaden, for useful discussions. This work was supported by NSF Grant No. PHY-9313781.

13. We are not aware of a general proof of the previous statement; it seems however plausible, given the variety of gauge fixing procedures for which it has already been verified. See also J. Papavassiliou and A. Sirlin, BNL preprint BNL-60241 (1994), to appear in Phys.Rev.D, where the application of the PT in the Standard Model quantized in the unitary gauge, gives rise to exactly the same renormalizable and g.i answer for the one-loop $W$ self-energy, as in the $R_\xi$ gauges.

14. As we will see shortly, this is not true in general for Green's functions involving off-shell external quarks.

15. The cancellation of $\xi_C$ from the S-matrix is automatic, due to current conservation.

16. This is indeed true for the simpler case, where only the gluon momentum is off-shell, with the two quarks on shell (see for example [3]).

17. Since constant terms are physically irrelevant, one cannot distinguish on physical grounds between the results obtained via the PT and those obtained in the BFM, for *arbitrary* values of $\xi_Q$.

18. In the context of the BFM, the gluon self-energy is known to comply with the running coupling up to at least two loops [9].

19. W. Fischler, Nucl. Phys. <u>B 129</u>, 157 (1977);
    A. Billoire, Phys. Lett. <u>92 B</u>, 343 (1980).

20. J. Papavassiliou, K. Philippides, and M. Schaden, in preparation.

## 2. Figure Captions

(i) **Figure 1 :** Graphs (a)-(d) are some of the contributions to the S-matrix $T$. Graphs (e), (f) and (g) are pinch parts, which, when added to the usual self-energy graphs (a), give rise to a gauge independent effective self-energy.

(ii) **Figure 2:** The general structure of the S-matrix elements used for the construction of gauge-independent three and four gluon vertices [(2a) and (2b), respectively], and some typical diagrams contributing vertex-like pinch contributions.

(iii) **Figure 3:** Graphs (3b) and (3c) are the gauge dependent parts arising from the conventional self-energy graph (3a); they cancel against pinch contributions, such as (3e), (3g) and (3i).

$$T(t,s,m_1,m_2) \;=\; \text{(a)} \;+\; \text{(b)} \;+\; \text{(c)} \;+\; \text{(d)}$$

$$\hat{T}_1(t) \;=\; \text{(a)} \;+\; \text{(e)} \;+\; \text{(f)} \;+\; \text{(g)}$$

Figure 1

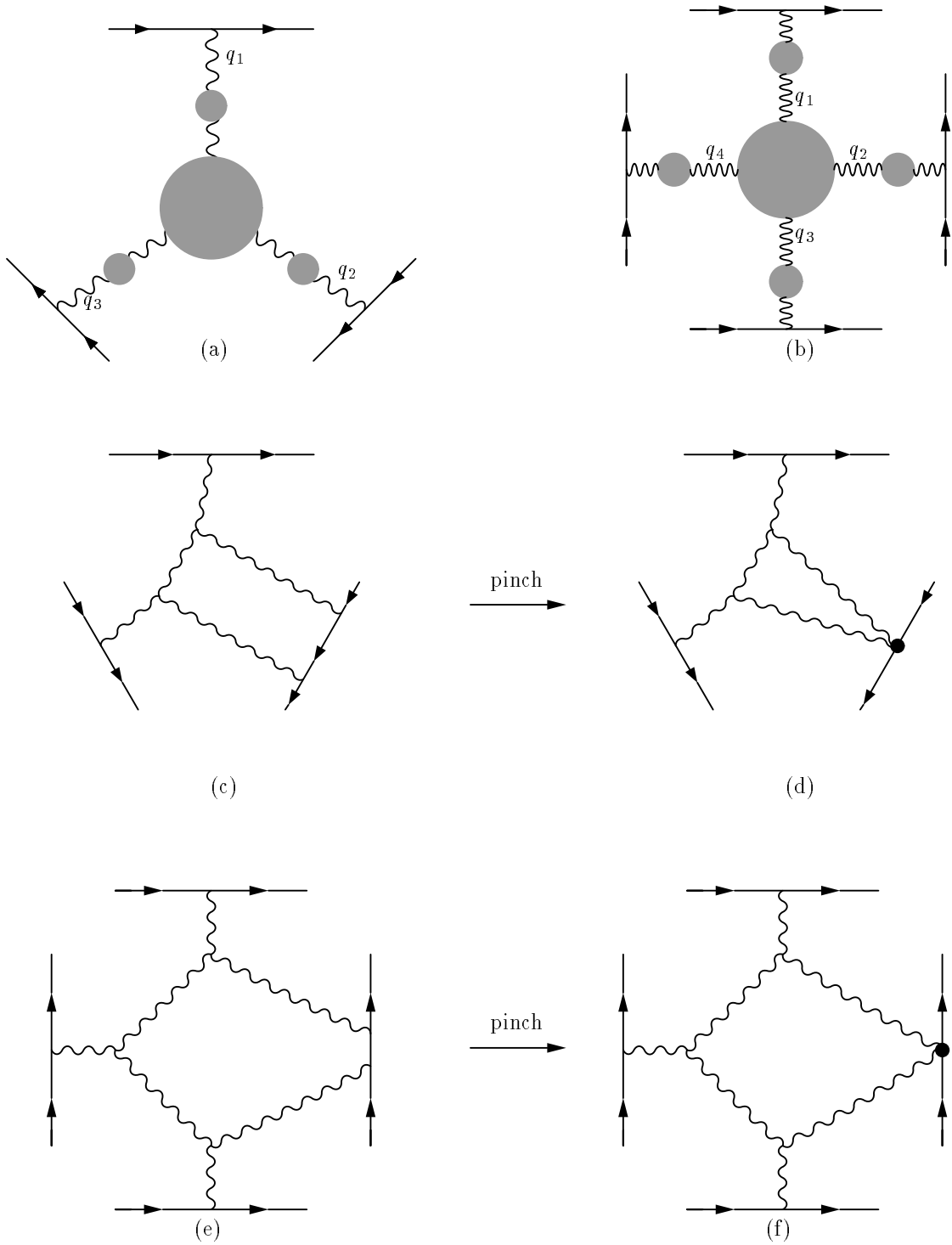

Figure 2

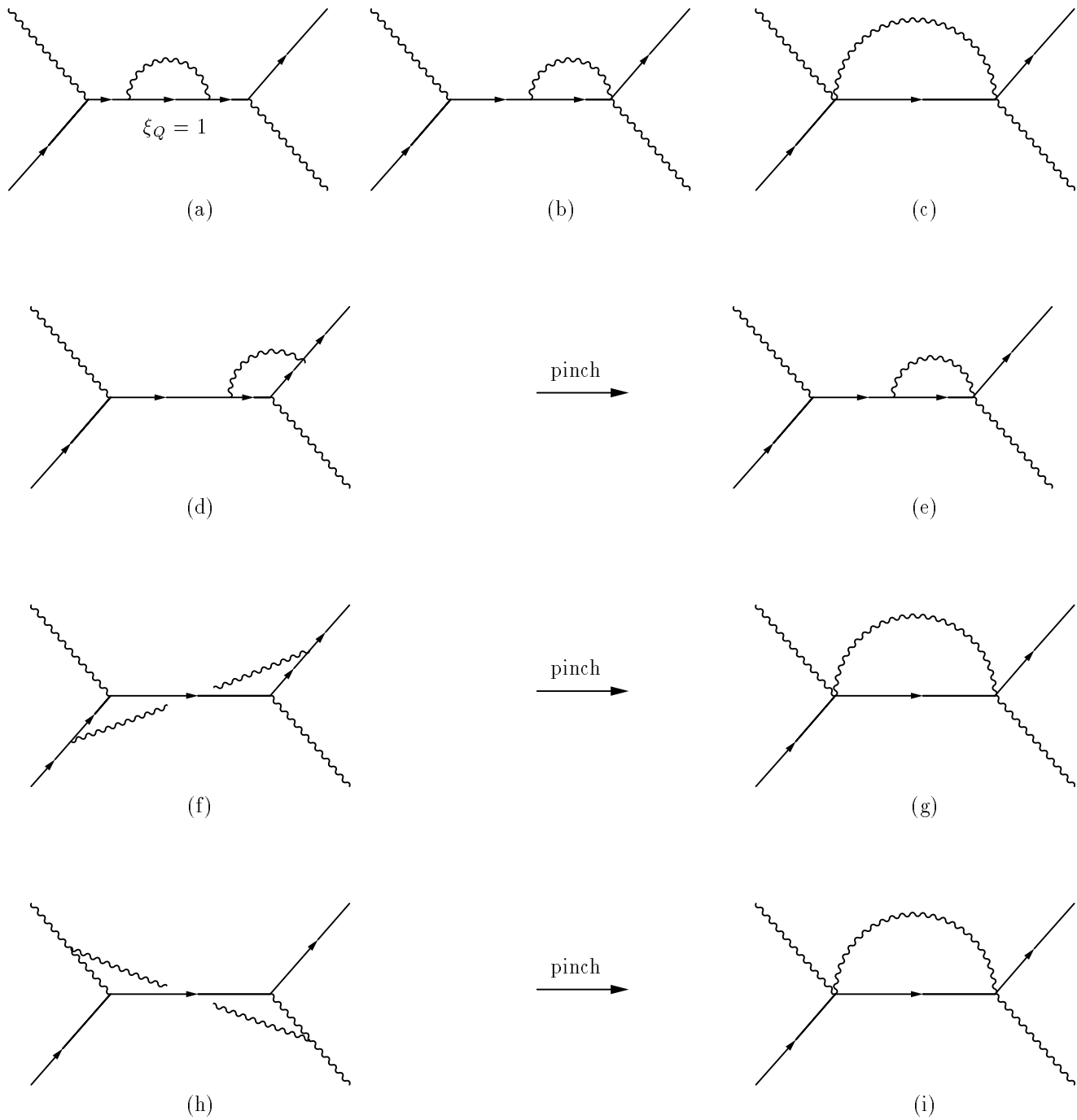

Figure 3